\documentclass[aps,prd,showkeys]{revtex4-1}
\usepackage{epsfig}
\usepackage{amssymb}
\usepackage{amsfonts}
\usepackage{epsfig}
\usepackage{bm,amsmath}
\usepackage{amsthm,amssymb}
\usepackage{graphicx}
\usepackage{dcolumn}
\usepackage{mathrsfs}
\usepackage{amsmath}
\usepackage{latexsym}
\usepackage{multirow}

\newcommand{\field}[1]{\mathbb{#1}}

\def\pd{\partial}
\def\G{\Gamma}
\def\f{\phi}
\def\m{\mu}
\def\n{\nu}
\def\mn{{\mu\nu}}
\def\nb{\nabla}
\def\dx{d^4x}
\def\l{\lambda}
\def\r{\rho}
\def\b{\beta}
\def\a{\alpha}
\def\Gt{{\tilde{G}}}
\def\lt{{\tilde{\l}}}
\def\gt{{\tilde{g}}}

\def\c{\chi}

\def\h{\eta}
\def\L{\Lambda}
\def\Lt{{\tilde{\L}}}
\def\Gb{{\bar{G}}}
\def\Lb{{\bar{\L}}}
\def\lb{{\bar{\l}}}
\def\y{\psi}
\def\k{\kappa}

\usepackage[usenames]{color}

\definecolor{Blue}{rgb}{0.0,0,1.0} 
\definecolor{Green}{rgb}{0.0,0.8,0.0} 
\definecolor{MyDarkBlue}{rgb}{0.1,0,0.55}

\def\al{\alpha} 
\def\be{\beta} 

\def\de{\delta}

\def\et{\eta}
\def\th{\theta}

\def\ka{\kappa}
\def\la{\lambda}

\def\si{\sigma}

\def\Ga{\Gamma}

\newcommand{\ben}{\begin{equation}}
\newcommand{\een}{\end{equation}}
\newcommand{\bea}{\begin{eqnarray}}
\newcommand{\eea}{\end{eqnarray}}
\newcommand{\ba}{\begin{array}}
\newcommand{\ea}{\end{array}}
\newcommand{\bit}{\begin{itemize}}
\newcommand{\eit}{\end{itemize}}

\def\pa{\partial}

\def\P{{\mathcal{P}}}
\def\R{{\mathcal{R}}}
\def\laplace{\triangle}

\def\bp{{\bf p}}

\def\bx{{\bf x}}

\def\half{\frac{1}{2}}

\def\mpl{m_{\rm Pl}}
\def\PS{\P}

\def\tpr{t_{\rm pr}}

\def\nuRG{\nu_{RG}}
\def\siRG{\si_{RG}}
\def\etaRG{\eta_{RG}}

\begin{document}

\title{
Renormalisation group improvement of scalar field inflation
}
\author{Adriano Contillo}
\affiliation{SISSA, 
via Bonomea 265, 34136, Trieste, Italy and INFN, sezione di Trieste}
\author{Mark Hindmarsh}
\affiliation{Department of Physics \& Astronomy,  University of Sussex, Brighton BN1 9QH, UK}
\author{Christoph Rahmede}
\affiliation{Department of Physics \& Astronomy, 
University of Sussex, Brighton BN1 9QH, UK}

\begin{abstract}
We study quantum corrections to Friedmann-Robertson-Walker cosmology with a scalar field under the assumption
 that the dynamics are subject to renormalisation group improvement.
 We use the Bianchi identity to relate the renormalisation group scale to the scale factor and obtain the improved cosmological evolution equations. We study the solutions of these equations in the renormalisation group fixed point regime, obtaining the time-dependence of the scalar field strength and the Hubble parameter in specific models with monomial and trinomial quartic scalar field potentials. We find that power-law inflation can be achieved in the renormalisation group fixed point regime with the trinomial potential, but not with the monomial one.  We study the transition to the quasi-classical regime, where the quantum corrections to the couplings become small, and find classical dynamics as an attractor solution for late times. We show that the solution found in the renormalisation group fixed point regime is also a cosmological fixed point in the autonomous phase space. We derive the power spectrum of cosmological perturbations and find that the scalar power spectrum is exactly scale-invariant and bounded up to arbitrarily small times, while the tensor perturbations are tilted as appropriate for the background power-law inflation. We specify conditions for the renormalisation group fixed point values of the couplings under which the amplitudes of the cosmological perturbations remain small.
\end{abstract}

\maketitle

\section{Introduction}
To describe the first instant of the evolution of the universe until the Planck-scale a theory of quantum gravity is needed. 
A viable proposal for such a theory is based on the assumption that the ultraviolet behaviour of the gravitational couplings is controlled by a renormalisation group fixed point \cite{Weinberg:1980gg}.
Then the metric field remains the fundamental carrier of the gravitational force and quantum corrections should take the form of a simple modification of  couplings whose scale-dependence is described by the Renormalisation Group (RG). 

Evidence that asymptotic safety could be realised comes from RG studies in the continuum \cite{Reuter:1996cp,Dou:1997fg,Souma:1999at,Lauscher:2001ya,Lauscher:2002sq,Reuter:2001ag,Litim:2003vp,Percacci:2003jz,Percacci:2005wu,Litim:2006dx,Niedermaier:2006wt,Niedermaier:2006ns,Percacci:2007sz,Litim:2008tt,Codello:2006in,Codello:2007bd,Codello:2008vh,Narain:2009fy,Narain:2009gb,Benedetti:2009gn,Machado:2007ea}, for reviews see  \cite{Litim:2006dx,Niedermaier:2006wt,Niedermaier:2006ns,Percacci:2007sz,Litim:2008tt},
and numerical studies on the lattice \cite{Ambjorn:2001cv,Hamber:1999nu,Hamber:2004ew}. RG effects from gravity would become important at high energies and thus also at the beginning of the Universe where the physics could be controlled by the RG fixed point.

This idea  has been explored in Einstein gravity  with an ideal fluid 
\cite{Bonanno:2001xi,Bonanno:2002zb,Contillo:2010ju,Koch:2010nn}, and in the context of $f(R)$ gravity \cite{Weinberg:2009wa, Tye:2010an,Bonanno:2010bt}. There is also the possibility of an infrared (IR) fixed point which can play a role in the observed acceleration of the Universe today \cite{Bonanno:2001hi,Bentivegna:2003rr,Reuter:2005kb,Babic:2004ev}.  Perturbative RG approaches to cosmology have  explored the running of the cosmological constant~ \cite{Shapiro:1999zt,Shapiro:2000dz,Shapiro:2008yu,Shapiro:2009dh} and of Newton's coupling \cite{Shapiro:2004ch,Bauer:2005rpa,Bauer:2005tu} and their implications for big bang nucleosynthesis, supernova observations, and deviations from standard cosmology \cite{Guberina:2002wt,Shapiro:2003kv,EspanaBonet:2003vk,Shapiro:2004ch}.

The RG scale parameter is assumed to be related to cosmological time
as RG effects are supposed to become significant at early times. For the exact form of such a relation there have been numerous proposals. Reuter et al.~\cite{Bonanno:2001xi,Bonanno:2001hi,Reuter:2005kb} chose the RG scale inversely proportional to cosmological time, and proportional to the Hubble scale. Other approaches considered as the relevant scale the fourth root of the energy density \cite{Guberina:2002wt}, or the cosmological event and particle horizons \cite{Bauer:2005rpa,Bauer:2005tu}. 
All of these approaches result in RG improved cosmological equations, with the potential to even generate entropy \cite{Bonanno:2007wg,Bonanno:2008xp}.  

In \cite{Reuter:2005kb,Hindmarsh:2011hx,Ahn:2011qt}, the RG scale was adjusted in such a way that the form of  the classical equations is unaltered. Preserving the Bianchi identity while taking into account the scale-dependence of the scalar-field couplings, requires  an evolution equation for the RG scale \cite{Hindmarsh:2011hx}. We will use the same consistency condition in this paper, so that the RG scale emerges dynamically. We will see that under certain circumstances the scalar field sets the relevant coarse-graining scale.

In Section \ref{s:Conservation}, we review how the requirement of preserving the Bianchi-identity with running couplings and no energy transfer to matter fixes the RG scale. In Section \ref{NGFP} we study the consequence of this relation in the RG fixed point regime for the  cases of monomial and quartic trinomial potential. We find the time-dependence of the scalar field strength and the Hubble parameter in the fixed point regime showing that monomial potential does not admit accelerated expansion whereas trinomial quartic potential does.  In section \ref{GFP} we study the quasi-classical regime where the quantum corrections to the running couplings are small and the beta functions can be linearised. In Section \ref{s:CosFixPoi} we study how the obtained results are related to fixed points of cosmological dynamics as studied in \cite{Hindmarsh:2011hx}. In Section \ref{seccomfluc}  we derive an expression for the power spectra generated by the quantum fluctuations around the RG improved background. 
We specify conditions on the fixed point values of the couplings  to fit the experimental requirement that cosmological perturbations be small.

\section{Renormalisation group improved stress-energy conservation }\label{s:Conservation}

As a first approach to the issue, we choose to consider an Einstein-Hilbert action minimally coupled to a scalar field, postponing the extension to non-minimal couplings to future work
\begin{equation}
 \G[g,\f]=\G_\text{EH}+\G_\f=\int\dx\sqrt{-g}\left[-\frac{R}{16\pi G}+\frac{1}{2}\nb_\m\f\nb^\m\f+V(\f)\right].
\end{equation}
We will discuss specific forms for the potential in later sections, but will keep it generic in this section.
At high energies, the couplings receive radiative corrections which can be described by renormalisation group running. We will assume that the dominant RG running in $\G_\f$ is contained in the potential $V$. The potential is assumed to take the form $V(\phi) = \sum_i\lambda_i\phi^i$, with $i$ a non-negative integer.

One can think of the renormalisation group (RG) as the generator of a one-parameter family of actions, with the parameter $k$ spanning over $\field{R}^+$. In physical situations, one has to choose a suitable value for this parameter. Deriving an expression for it is the main goal of the present discussion, and we are following here the same approach as \cite{Reuter:2005kb,Hindmarsh:2011hx}.
In the neighbourhood of
a given event $A$, one can define an action $\G_A\equiv\G[g,\f;k_A]$, where the value of $k_A$ will not be specified for the time being. The resulting field equations
\begin{eqnarray}\label{Einsteineqs}
 G^\mn &=& 8\pi G_AT_A^\mn=8\pi G_A\left(\nb^\m\f\nb^\n\f-\frac{1}{2}g^\mn\nb_\r\f\nb^\r\f-g^\mn V_A(\f)\right) \\
\label{e:sfEqn}
  \Box\f &=& V'_A(\phi) 
\end{eqnarray}
obtained by varying $\G_A$ with respect to the metric $g$ and the scalar field $\phi$, are valid in a neighbourhood of $A$. Moreover, as $\G_A$ is a diffeomorphism invariant action, both the resulting Einstein tensor $G^\mn$ and the stress-energy tensor $T_A^\mn$ (where the subscript $A$ indicates that the couplings are evaluated at $k_A$) are covariantly conserved in the neighbourhood of  $A$.

In a close but distinct event $B$, same considerations as for $A$ apply, but we let $k_B\neq k_A$. In our picture, each action has different but constant couplings, and each set of field equations is the Einstein one. Iterating the procedure for each event $x$ and reducing the space-time patches to infinitesimal size, we obtain solutions of the Einstein equations in each patch which depend on a different $k$. The resulting one-parameter set of solutions can then be described by making $k$ a function of the space-time coordinates. Doing so does not affect the RG procedure itself.

The approach chosen here, where running couplings are inserted into the standard equations of motion, has been dubbed  \textit{restricted}  improvement in ~\cite{Bonanno:2010bt}.
A different approach, which was called \textit{extended} improvement in~\cite{Bonanno:2010bt}, was studied previously also  in~\cite{Reuter:2003ca}.  There the couplings are treated as some sort of external fields $G(x)$ in the action, 
which results in field equations with a source term proportional to $\nb_{\!\m} G$. This approach will not be followed here. 

We split the covariant derivatives of the Einstein and stress-energy tensor into a part only acting on the fields and one only acting on the couplings
\begin{eqnarray}\nonumber
 \nb_\m\left(\frac{G^\mn}{8\pi G}\right)&=&\frac{1}{8\pi G}\nb_\m G^\mn-\frac{G^\mn}{8\pi G^2}\nb_\m G\\
 \nb_\m T^\mn&=&\nb_\m\left.T^\mn\right|_\l-\left. \nb^\n V(\phi)\right|_{\phi}
 \end{eqnarray}
with the subscripts $\l$, $\phi$ meaning that the couplings or the scalar field are kept constant.

In the restricted improvement picture, in the limit that $k$ is a continuous function of $x$, the stress-energy tensor remains conserved at fixed $k$,  as diffeomorphism invariance of \textit{each} action is preserved in the limiting process.
We must also satisfy the Bianchi identity.
Hence 
\begin{equation}\label{partcons}
 \nb_\m G^\mn=0,\qquad \qquad\nb_\m\left.T^\mn\right|_\l=0,
\end{equation}
and the latter is obviously consistent with the scalar field equation  (\ref{e:sfEqn}).
The derivation of the field Eqs. (\ref{Einsteineqs}) guarantees the total conservation
\begin{equation}\label{totcons}
 \nb_\m G^\mn=8\pi\nb_\m(G\,T^\mn)
\end{equation}
so that substituting (\ref{partcons}) in (\ref{totcons}) gives 
\begin{eqnarray}\label{constraint}
 \nb_\m G T^\mn 
  -G
\left. \nb^\n V(\phi)\right|_\phi
  &=&
 0
 \;.
\end{eqnarray}
One can think of Eqs. (\ref{Einsteineqs}) and (\ref{e:sfEqn}) as a family of two equations labelled by the value of $k(x)$, each one valid only in a neighbourhood of $x$, with (\ref{constraint}) the condition required to connect neighbouring spacetime points.
Moreover, it is straightforward to notice that, in a standard framework with no RG improvement, Eq. (\ref{constraint}) is trivially satisfied. This means that in the present case it represents a genuine new constraint that can be used to define the function $k(x)$.

It is convenient to rewrite the equations in dimensionless form by the introduction of dimensionless couplings, \emph{i.e.} measured in units of the RG scale $k$,
\begin{equation}
 G(k)=k^{-2}\Gt(k)\qquad;\qquad 
 \qquad \l_i(k)=k^{4-i}\lt_i(k)
\end{equation}
and their covariant derivatives as
\begin{eqnarray}
 \nb_\m G&=&\nb_\m k\frac{dG}{dk}=\frac{\nb_\m k}{k^3}\left(\b_\Gt-2\Gt\right)
 \\
  \nb_\m \l_i&=&\nb_\m k\frac{d\l_i}{dk}=\frac{\nb_\m k}{k^{i-3}}\left(\b_{\lt_i}+(4-i)\lt_i\right).
\end{eqnarray}
Here we introduce the $\be$-functions for the dimensionless couplings $\b_\Gt$, $\b_{\lt_i}$, defined in the usual way.
The system (\ref{Einsteineqs},\ref{e:sfEqn},\ref{constraint}) reads
\begin{eqnarray} \label{Einstein2}
G^\mn &=& 8\pi G T^\mn 
\\
\label{field2}
 \Box\f &=& V' \\
\label{constraint2}
( \nb_\m \ln k )T^\mn \et_{\rm RG}
 &=& (\nb^\n\ln k) V \nu_{\rm RG}
\end{eqnarray}
where we used the definitions
\begin{eqnarray}
\label{etadefinition}
\eta_{RG}&=&\frac{\pd\ln G}{\pd\ln k} =\frac{\b_\Gt}{\Gt}-2 \\
\label{nudefinition}
\nu_{RG}&=&\frac{\pd\ln V}{\pd\ln k} = \frac{1}{V}\sum_i\left(\b_{\lt_i}+(4-i)\lt_i\right)k^{4-i}\phi^i
\end{eqnarray}
In order to write down (\ref{constraint2}) we assume that the energy-momentum tensor does not acquire a further anomalous dimension, which  by the equivalence principle would not be expected in the quasi-classical regime as it is a conserved current. We also assume that the most important running of the energy-momentum tensor is contained in the scalar potential and characterised by the parameter $\nu_{RG}$.
Wavefunction renormalisation would induce a term $\sim\gamma_{RG}\,Z\,(\partial\phi)^2$ in Eq. (\ref{Einstein2}), where $\gamma_{RG}$ is the anomalous dimension of the scalar field.  As the wavefunction renormalisation comes in only at two loops for a scalar, the dominant renormalisation effects are in the constant and quadratic terms of the scalar potential. The quasi-classical regime where $\eta_{RG}\to 0$ is discussed in section \ref{GFP}. There we will see that vanishing $\eta_{RG}$ entails also vanishing $\nu_{RG}$.

Eq.\ (\ref{constraint2}) shows that $ \nb_\m \ln k $ is an eigenvector of the stress-energy tensor, with eigenvalue $V\left({\nu_{RG}}/{\eta_{RG}}\right)$.  If $\nb_\m \ln k $ is timelike, it must therefore be proportional to the fluid velocity four-vector $u^\mu$, whose eigenvalue is minus the energy density $\rho$.  Hence \cite{Hindmarsh:2011hx}
\ben\label{evfluid}
\frac{V}{\rho} =  - \frac{\eta_{RG}}{\nu_{RG}}.
\een
 An important corollary is that in a coordinate system which is comoving with the fluid (i.e.\ for which $T^{0i} = 0$), $k$ must be a function of time $t$ only.

 \section{Cosmology in the fixed point regime}\label{NGFP}

In this section and the following ones, we apply the previous considerations within the context of standard cosmology on a flat FRW-background. We correct the equations of motion of cosmology by running couplings, assuming that the running can be translated into a time-dependence.

At very early times, asymptotic safety would lead to the RG dependence of the couplings being controlled by the RG fixed point, making the couplings scale according to their mass dimension. On dimensional grounds we will look for solutions in which the RG parameter $k$ is inversely proportional to cosmological time $t$.

At the RG fixed point, the dimensionless couplings approach constant values and the beta functions vanish,
\begin{equation}\label{FPcouplings}
 \gt(k)\simeq\gt^\ast\qquad\Rightarrow\qquad\b_\gt\simeq0
\end{equation}
with $\gt=\{\Gt,\lt_i\}$. As a result, the system (\ref{Einstein2},\ref{field2},\ref{constraint2}) can be rewritten in the form
\begin{eqnarray}
 H^2&=&\frac{8\pi G}{3}\left(\frac{1}{2}\dot{\f}^2+V
 \right)\label{friedmann}\\
\dot{H}&=&-4\pi G\dot{\f}^2\label{Hdot}   \\ 
\ddot{\f}&=&-3H\dot{\f}-V'\label{field3}\\
\dot{\f}^2&=&
 -2\left(1+\frac{\nu_{RG}}{\eta_{RG}}\right) V\label{constraint3}
\end{eqnarray}  
where $H=\dot{a}/a$ is the Hubble parameter.
In the case that the bracket in the last of the four equations becomes zero,
 the scalar field and the Hubble parameter become constant giving a de Sitter solution with inflationary expansion provided $V'=0$. In the following we will assume that it is different from zero.
Then, using Eq. (\ref{constraint3}) and the resulting relation between $H$ and $V$ one obtains the equations
\begin{eqnarray}
&\ddot{\phi}=-\sqrt{48\pi G\frac{\nu_{RG}}{\eta_{RG}}\left(1+\frac{\nu_{RG}}{\eta_{RG}}\right)}V-V'&\label{phidecoupled}\\
&-\frac{\dot{H}}{H^2}=\frac{1}{\alpha}&\label{Hdecoupled}
\end{eqnarray}
where 
\begin{equation}\label{alphasol}
\alpha=\frac{1}{3\left(1+\frac{\eta_{RG}}{\nu_{RG}}\right)}\ .
\end{equation}
Thus the value of the Hubble rate depends only on the dimensionless RG parameters $\nu_{RG}$ and $\eta_{RG}$. In general they will depend on the dimensionless ratio $\tilde\phi=\phi/k$. When this ratio becomes constant, power-law solutions for $H$ are obtained. 
The condition which gives  inflationary expansion is 
\begin{equation}
-\frac{\dot{H}}{H^2}<1.
\end{equation}
This is fulfilled for 
\begin{equation}\label{FPinflacond}
-\frac{\eta_{RG}}{\nu_{RG}}>\frac{2}{3}\ .
\end{equation}
In the fixed point regime, this condition can be met for any potential as long as 
${\nu_{RG}}<3$.

The system (\ref{phidecoupled},\ref{Hdecoupled}) can be solved, at least numerically, for any form of the field potential, treating the fixed point values of the couplings as given quantities. In some special cases it is possible to derive an analytic solution that allows us to treat the couplings as free parameters. 
Including only the cosmological constant in the potential, $V(\f)=\Lambda/(8\pi G)$, at a fixed point the RG parameters become 
\begin{equation}
\eta_{RG}=-2\ ;\qquad \nu_{RG}=4\ .
\end{equation}
Thus, at the corresponding  RG fixed point one has $\dot{\f}^2/2=V(\f)$ indicating equality between kinetic and potential energy of the scalar field as found in \cite{Bonanno:2001xi,Bonanno:2002zb}. Note that this is also the case at the so-called Gaussian matter fixed points, where the dimensionless scalar field couplings vanish. We study here the further interesting cases of a monomial potential and a quartic trinomial potential which allow analytic solutions.

In general, we would expect to align the RG scale $k$ with the one of the two physical scales in the problem, the Hubble parameter $H$ or the scalar field expectation value $\phi$.  Most works with RG corrections in cosmology assume that $H$ sets the relevant coarse-graining scale for fluctuations around the spatially homogeneous background, and indeed if there is no special tuning, we would expect to be able to find a solution with $k \sim H \sim \phi$ at a fixed point.  The assumed spatial homogeneity of an FRW universe means that it is consistent to take $k$ to be a function of time only, as in (\ref{evfluid}).

A possible ansatz to solve Eqs. (\ref{friedmann},\ref{Hdot},\ref{field3},\ref{constraint3}) is therefore to assume that the Hubble rate, field strength and RG parameter all scale inversely proportional to time,
\begin{equation}\label{Ansatzpropt}
H=\frac{\alpha}{t}\ ; \qquad \phi=\frac{\varphi}{t}\ ; \qquad k=\frac{\chi}{t}
\end{equation}
where $\varphi$, $\c$ are constants to be determined and $\alpha$ is given by Eq. (\ref{alphasol}).

We will find in Section \ref{seccomfluc} that if the fixed point values of the dimensionless couplings satisfy certain special conditions, the expansion rate $H$ can be much smaller than $k$ and $\phi$.  In an inflating solution near the Gaussian fixed point (Section \ref{GFP}) we will find that it is generally true that $H \ll k \sim \phi$, i.e.\ that consistency with the Bianchi identity (\ref{constraint2}) forces one to coarse-grain at the scale set by the scalar field.


Inserting the ansatz (\ref{Ansatzpropt}) into Eqs. (\ref{friedmann},\ref{Hdot},\ref{field3},\ref{constraint3}) leads to
\begin{eqnarray}
\alpha^2&=&\frac{8\pi\tilde G}{3}\left(\frac{1}{2}{\tilde\phi}^2+\chi^2\tilde V\right)\\
\a&=&4\pi\Gt\, {\tilde\phi}^2
\label{mufromalpha}\\
2 {\tilde\phi}&=&3\a {\tilde\phi}-\c^2{\tilde V'}\\
 {\tilde\phi}^2&=&-2\left(1+\frac{\n_{RG}}{\h_{RG}}\right)\c^2\tilde V
\label{chisol}
\end{eqnarray} 
where $\tilde\phi = \phi/k$, ${\tilde V}( {\tilde\phi})=
k^{-4} V(\phi)$ and 
${\tilde V'}( {\tilde\phi})=
k^{-3} V'(\phi)$. One can check that these equations give consistent solutions for the three parameters $\alpha$, $\chi$ and $\tilde\phi$.  
Additionally one obtains
\begin{equation}
\varphi^2=-\frac{ {\tilde\phi}^4}{2\left(1+\frac{\n_{RG}}{\h_{RG}}\right)\tilde V}\label{jsol}
\end{equation}
 If one chooses a specific potential, the solution for $ {\tilde\phi}$ and $\chi$ can be completed.  We will discuss the cases of monomial and trinomial potential in the next two subsections.

\subsection{Monomial potential}\label{NGFPmon}

As a simple test case, we consider
a monomial potential of the form
\begin{equation} 
 V(\phi)=\lambda_n\phi^n,
 \end{equation}
where $n$ is an integer.  In the fixed point regime $\nu_{RG}=4-n$, as explained in the Appendix, giving (here and in the following section, all couplings assume their fixed point values)
\begin{eqnarray}
\a&=&\frac{4-n}{3\,(2-n)}\\
\c^2&=&\frac{1}{(2-n){\tilde\lambda}_n}\left(\frac{4-n}{12(2-n)\pi\Gt}\right)^{\frac{2-n}{2}} \label{e:ChiEqn}\\
\varphi^2&=&\frac{1}{(2-n){\tilde\lambda}_n}\left(\frac{4-n}{12(2-n)\pi\Gt}\right)^{\frac{4-n}{2}}\ . \label{e:VarPhiEqn}
\end{eqnarray}
In this case, $\alpha$ is constant and hence there is 
power-law expansion with $a\propto t^{\a}$. An expanding universe requires positive $H$.  This can only be achieved if $\a>0$, and, in the fixed point regime where $\h_{RG}=-2$, this needs either $\n_{RG}<2$ or $\n_{RG}>4$.
We exclude the case $n=2$ because it is inconsistent with (\ref{Ansatzpropt}).
The case $n=4$ would give $H=0$ and thus not lead to a realistic scenario.
Eq. (\ref{FPinflacond}) implies that inflationary expansion cannot be obtained in the fixed point regime with an even monomial potential.
Inflation may however be obtained if one includes two or more terms in the potential.

\subsection{Quartic potential}\label{secquartic}

A more realistic model arises from the non-singular and symmetric trinomial potential including only couplings with positive or zero mass dimension
\begin{equation}
 V(\f)=\l_0+\l_2\,\f^2+\l_4\,\f^4
\end{equation}
We can still derive an analytic solution of the system (\ref{friedmann},\ref{Hdot},\ref{field3},\ref{constraint3}). This happens because in this particular case (\ref{constraint3}) reads
\begin{equation}
 \dot{\f}^2=2\lt_0\,k^4-2\lt_4\,\f^4
\end{equation}
and $k$ can be extracted and plugged into (\ref{field3}) \footnote{Note that in the case without cosmological constant, $\l_0=0$, a solution can be obtained if $\l_4$ is negative. Then one obtains $\phi=\frac{\phi_0}{1-\sqrt{-2 \l_4}\phi_0 t}$, where $\phi_0$ is some initial value.}.
From Eqs. (\ref{alphasol},\ref{chisol},\ref{jsol}) one obtains 
\begin{equation}
\label{e:AlpChiPhi}
\a=\frac{2\lt_0+\lt_2 {\tilde\phi}^2}{3(\lt_0-\lt_4 {\tilde\phi}^4)}\ ;\qquad
\c^2=\frac{ {\tilde\phi}^2}{2(\lt_0-\lt_4 {\tilde\phi}^4)}\ ;\qquad
\varphi^2=\frac{ {\tilde\phi}^4}{2(\lt_0-\lt_4 {\tilde\phi}^4)}
\end{equation}
whereas Eq. (\ref{mufromalpha}) gives $\tilde\phi$ as a solution to the equation
\begin{equation}\label{phiequation}
\lt_0-\lt_4 {\tilde\phi}^4=\frac{1}{12\pi\Gt}\left(\frac{2\lt_0}{ {\tilde\phi}^2}+\lt_2\right)\ .
\end{equation}
This equation can be solved analytically with Cardano's formula, but we do not find it useful to be displayed here. We will need this result however only in Section \ref{seccomfluc} in the case where both sides become small. 

\begin{figure}[!ht]
\begin{center}
 \includegraphics[width=10cm,height=6cm]{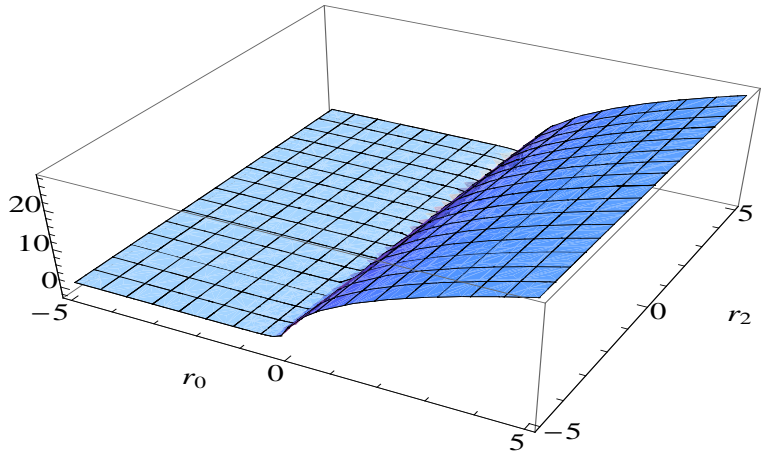}
\end{center}
\caption{Numerical value of $\a$, for $r_0$ and $r_2$ spanning the range $\{-5,5\}$.}
\label{tplot}
 \end{figure}

The parameters $\alpha$ and $\chi$ depend only on the combinations of couplings
 $r_0=\lt_0/\lt_4$ and $r_2=\lt_2/\lt_4$. With respect to these two parameters, we show in Figure \ref{tplot} the value of $\a$. It can be seen that for sufficiently positive $r_0$ it is greater than one, thus describing a phase of power law inflation.

\section{The quasi-classical regime}\label{GFP}

The picture we are left with from the previous section is that the universe undergoes an inflationary phase driven by the UV fixed point, where the quantities $H$, $\phi$ and $k$ decrease inversely proportional to cosmological time. In particular, the energy scale $k(t)$ goes down until it reaches the value at which the RG trajectory leaves the UV fixed point and starts to roll towards the infrared.

Among the infinite number of trajectories originating from the UV fixed point, the physical one must be chosen in such a way to provide a long-lasting phase of (almost) classical cosmological evolution. One can achieve this by restricting oneself to the bunch of trajectories that lie very close to the separatrix, defined as the trajectory connecting the UV fixed point and the Gaussian fixed point, where $\Gt=\lt_i=0$.

After having left the UV fixed point, such a trajectory rapidly falls downwards until it borders on the Gaussian fixed point. During this transition phase, the universe expands and cools down, quantum effects become more and more negligible and the transition to classicality takes place. After a finite amount of (RG) time the trajectory departs from the linear regime and tends towards its deep IR regime, that can possibly be located at another fixed point, but this issue is beyond the purposes of the present analysis.

Once the trajectory has left the fixed point, the approximation (\ref{FPcouplings}) ceases to hold, but because of the small values of the couplings, the $\b$-functions (as calculated in~\cite{Narain:2009fy}) can be linearised and read to first order in $\tilde G$ and the matter couplings
\begin{eqnarray}\label{linbetas}\nonumber
 \b_\Gt&=&2\,\Gt\\ \nonumber
 \b_\Lt&=&\frac{3\,\Gt}{4\pi}-2\,\Lt\\
 \b_{\lt_2}&=&-2\,\lt_2-\frac{3\,\lt_4}{8\pi^2}\\ \nonumber
 \b_{\lt_4}&=&0
\end{eqnarray}
where we made use of the usual definition $\Lt=8\pi\Gt\,\lt_0$. Obviously, a less trivial evolution can be obtained by including higher order terms in (\ref{linbetas}), but at least in the vicinity of the Gaussian fixed point these contributions will be negligible. The linearised flow (\ref{linbetas}) can be easily integrated, and the final $k$-dependence of the dimensionful couplings turns out to be
\begin{eqnarray}\label{flux}\nonumber
 G(k)&=&\Gb\\ \nonumber
 \L(k)&=&\Lb+\frac{3}{16\pi}\,\Gb\,k^4\\
 \l_2(k)&=&\lb_2-\frac{3}{16\pi^2}\,\lb_4\,k^2\\ \nonumber
 \l_4(k)&=&\lb_4
\end{eqnarray}
where the bars indicate the asymptotic values for very small $k$. From these equations one obtains $\eta_{RG}=0$. An overall remark that is worth making is that, for sufficiently small $k$, the quantum corrections to the flux (\ref{flux}) can be neglected and the dynamics become completely classical.

The flow (\ref{flux}) can be inserted into the constraint Eq. (\ref{constraint}) giving the fairly simple result
\begin{equation}
\nabla_{\mu}\ln k\; V\;\nu_{RG}=0
\end{equation}
that implies either the trivial case of constant $k$,  or vanishing potential, or $\nu_{RG}=0$. Here we will assume the latter condition, so that using the formula for $\nuRG$ for the case of the quartic potential in the Appendix and applying again (\ref{flux}), we find 
\begin{equation}\label{kvsf}
 k(t)=2\sqrt{\lb_4}\,\f(t).
\end{equation}
This can be plugged together with (\ref{flux}) into (\ref{field3}), giving 
\begin{equation}\label{linfield}
 \ddot{\f}+2\sqrt{6\pi\Gb\left(\frac{1}{2}\,\dot{\f}^2+\frac{\Lb}{8\pi\Gb}+\lb_2\,\f^2+\left(1-\frac{3\lb_4}{8\pi^2}\right)\lb_4\,\f^4\right)}\;\dot{\f}\,=-2\left(\lb_2+2\,\left(1-\frac{3\lb_4}{8\pi^2}\right)\lb_4\,\f^2\right)\,\f\;.
\end{equation}
This equation is again the Klein-Gordon equation, with modified but constant couplings. Nonetheless, it contains the identification (\ref{kvsf}), so that it can be thought of as an ``effective'' equation describing an intermediate phase located between the UV fixed point and the fully classical regime, described by Eq. (\ref{flux}) at $k=0$.

Eq. (\ref{linfield}) can be studied by means of the phase diagram method. We define $\y(\f)\equiv\dot{\f}$, so that $\ddot{\f}=\y\,\y'$ (the prime denoting the derivative with respect to $\f$) and (\ref{linfield}) can be rewritten as
\begin{equation}\label{psfield}
 \y'+2\sqrt{6\pi\Gb\left(\frac{1}{2}\,\y^2+\frac{\Lb}{8\pi\Gb}+\lb_2\,\f^2+\left(1-\frac{3\lb_4}{8\pi^2}\right)\lb_4\,\f^4\right)}\,
 =-2\left(\lb_2+2\,\left(1-\frac{3\lb_4}{8\pi^2}\right)\lb_4\,\f^2\right)\,\frac{\f}{\y}\;.
\end{equation}
Following the approach of~\cite{Mukhanov:2005sc}, we separate the phase space into kinetic and potential term dominated         
regions  ($|\y|\gtrless\f^2$) 
and study Eq. (\ref{psfield}) in both regimes. Notice        
that in these variables the previous fixed point phase is described by the parabola           
\begin{equation}                                                                              
\y=-\frac{\f^2}{\varphi}                                                                           
\end{equation}                                                                                
so that, when the RG trajectory leaves the fixed point, it is sufficient to restrict          
ourselves to the lower right quadrant ($\f>0$ and $\y<0$). Starting from the region           
$|\y|\gg\f^2$, Eq. (\ref{psfield}) admits the solution                                        
\begin{equation}                                                                              
\y(\f)=\y_0\,e^{-\sqrt{12\pi\Gb}\,\f}                                                          
\end{equation}                                                                                
that describes an exponential fall towards the region where $|\y|\ll\f^2$. Here the           
attractor solution identified by $\y'(\f)\approx0$ asymptotes towards the straight line   $\y(\f)
=-A\,\f$, where    
\begin{equation}                                                                              
A
=\sqrt{\frac{2}{3\pi\Gb}\left(1-\frac{3\,\lb_4}{8\pi^2}\right)\,\lb_4}\;    
.                                                                                             
\end{equation}                                                                                                                                                     
Trajectories originating from any point of the phase space are then forced over the           
attractor and follow it towards smaller values of $\f$. 

\begin{figure}[!h]
\begin{center}
 \includegraphics[width=12cm,height=4cm]{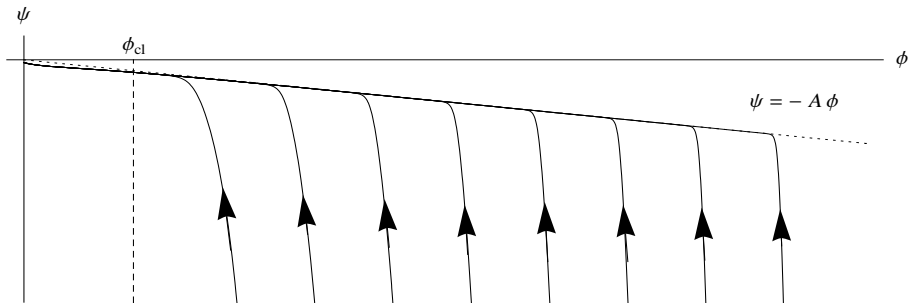}
\end{center}
\caption{Phase space portrait of the quasi-classical evolution of the scalar field $\f$. Dotted line represents the (approximated) attractor. The classical regime applies to the area on the left of the vertical line at $\f=\f_\textrm{cl}$.}
\label{psplot}
 \end{figure}

If we plug the condition              
(\ref{kvsf}) into (\ref{flux}), the beta functions for $\Lambda$ and $\l_2$ become            
$\phi$-dependent. We can thus determine for which field value the $\phi$-dependent part       
becomes negligible with respect to the asymptotic value. Such value, here named             
$\f_\textrm{cl}$, is the smaller value of  $\f_2$ and $\f_0$, where                
\begin{equation}                                                                              
\f_2=\left(\frac{4\pi^2}{3}\,\frac{\lb_2}{\lb_4^2}\right)^{1/2}                               
\ ; \qquad                                                                         
\f_0=\left(\frac{\pi}{3}\,\frac{\Lb}{\Gb\,\lb_4^2}\right)^{1/4}\ .                            
\end{equation}       
$\f_\textrm{cl}$ marks the         
scale at which RG cosmology becomes completely classical, and can be studied within the       
standard cosmological framework. The entire phase space portrait is shown in Figure           
\ref{psplot}.

\section{Autonomous system analysis}\label{s:CosFixPoi}

For a general phase space analysis, it is convenient to rewrite the equations of cosmological dynamics in autonomous form 
by introducing the dimensionless parameters
\begin{equation}
 x=\frac{\k\dot{\f}}{\sqrt{6}H}\; ;\qquad y=\frac{\k\sqrt{V}}{\sqrt{3}H}\; ;\qquad z=\frac{V'}{\k V}\; ;\qquad \eta=\frac{V''}{\kappa^2 V}\; ;\qquad N=-\ln a
\end{equation}
with $\k=\sqrt{8\pi G}$, see e.g.~\cite{Copeland:1997et,Copeland:2006wr,Hindmarsh:2011hx}.
In these variables the Friedmann equation turns into
\begin{equation}
 x^2+y^2=1
\end{equation}
confining the motion to an upper half-circle for an expanding universe (which has positive $y$).
The fraction of the total energy density carried by the kinetic and potential terms of the scalar field are 
$x^2$ and $y^2$. The variable $z$ is function of $\phi$ and the couplings, and can be used to recover the field value, and the function $\eta(z)$ encodes the potential $V(\phi)$.  With a specific $\eta(z)$ the equations of cosmological dynamics are represented in autonomous closed form in terms of dimensionless variables and a dimensionless evolution parameter $N$, the number of e-foldings. The interesting point is that the equations show fixed points where the equation of state parameter of the scalar field, which can be written as $2x^2/(x^2+y^2)$, becomes constant. That allows to trivially integrate to obtain the time dependence of the scaling factor. The specific properties (existence, attractivity) of the fixed points depend on the shape of the potential. Typical fixed points show e.g. domination by the kinetic term ($x=1$) or, during the slow-roll regime, by the potential term $y= 1$.

The modification of the fixed points of the equations of cosmological dynamics by RG effects constrained by Eq. (\ref{constraint}) has been studied in~\cite{Hindmarsh:2011hx}. 
There, correction terms were obtained which are parameterized by the RG parameters $\nu_{RG}$ and $\eta_{RG}$ as defined in Eqs. (\ref{etadefinition},\ref{nudefinition}), and a further parameter
\begin{equation}\label{sigmadefinition}                                                                                                                                                           
\sigma_{RG} = \frac{\partial \ln V'}{\partial \ln k}\ .                                                                                                   
\end{equation}        
Then the equations of cosmological dynamics can be written as (we do not take any ideal fluid components into account here)
\begin{eqnarray} 
\frac{{d}x}{{d}N}&=&
3x(1-x^2)+\sqrt{\frac{3}{2}}y^2z 
+ \frac{1}{2} x \eta_{RG}  \frac{d \ln k}{dN} , \label{eq:XeqnRG}\\
\frac{{d}y}{{d}N}&=&
-\sqrt{\frac{3}{2}}x yz - 3 x^2 y 
+  \frac{1}{2} y (\eta_{RG} + \nu_{RG})  \frac{d \ln k}{dN} \ ,\label{eq:YeqnRG}\\
\frac{{d}z}{{d}N}&=&-\sqrt{6}\,x(\eta(z)-z^2) + z\left(-\frac{1}{2} \eta_{RG} - \nu_{RG} + \sigma_{RG}\right)  \frac{d \ln k}{dN}\ , \label{eq:ZeqnRG}
\end{eqnarray}
and the constraint Eq. (\ref{constraint}) becomes
\begin{equation}
\eta_{RG}(k) + y^2\, \nu_{RG}(k,z) = 0\,. \label{eq:BiaCon}
\end{equation}
With the abbreviation
\begin{equation} \label{eq:RGalOmDef}
\alpha_{RG} = \frac{1}{2}\left[ \eta_{RG} + \nu_{RG} - \frac{\partial}{\partial \ln k} \ln \left(-\frac{\eta_{RG}}{\nu_{RG}}\right)\right]\, ,
\end{equation}
one finds
\begin{equation}
\frac{d \ln k}{dN}
 =\frac{1}{\alpha_{RG}}\left[\frac{\sigma_{RG}}{\nu_{RG}}\sqrt{\frac{3}{2}}x z + 3 x^2  
   \right]\, \label{eq:TeqnRG}\ .
\end{equation}
Without assuming any specific functional form for $k(t)$ and the potential, the Bianchi constraint leads immediately to 
\begin{equation}
x=\pm\sqrt{1+\frac{\eta_{RG}}{\nu_{RG}}}\ ;\qquad y=\sqrt{-\frac{\eta_{RG}}{\nu_{RG}}}
\end{equation}
We note that $ x = \pm\sqrt{{1}/{3\alpha}}$, where $\al$ was defined in Section  \ref{NGFP} as $tH$.
The Hubble parameter can be recovered by integrating 
\ben
\frac{d \ln H}{dN} = 3x^2,
\label{e:HubEqn}
\een

\subsection{The RG fixed point regime}

The equations (\ref{eq:XeqnRG},\ref{eq:YeqnRG},\ref{eq:ZeqnRG}) can show fixed points where $d\ln k/dN$ becomes either zero or where it approaches a constant value different from zero. In~\cite{Hindmarsh:2011hx} the first case was called a freeze-in fixed point, in which the evolution of $k$ with $N$ comes to a halt at some point. The second case gives instead rise to what was called simultaneous fixed points, in which the fixed points of  cosmological dynamics and the RG fixed points are reached together.  
The solution found in Section \ref{NGFP} is a simultaneous fixed point of a particular kind, in which $H \propto k \propto \phi$.  We denote this a scaling simultaneous fixed point.

Recalling (\ref{e:HubEqn}),  
we see that at a scaling simultaneous fixed point the conditions $H \propto k$  and $\phi \propto k$ imply that 
\bea
3x^2 &=& \frac{1}{\alpha_{RG}}\left[\frac{\sigma_{RG}}{\nu_{RG}}\sqrt{\frac{3}{2}}x z + 3 x^2  
  \right],   \label{e:SSFPconA} \\
   \frac{d \tilde\phi}{dN} &=& 0.
  \label{e:SSFPconB}
\eea
It is not at all obvious that the first of these equations is satisfied, and we proceed to demonstrate that it is consistent.

The  condition (\ref{e:SSFPconB}), coupled with $d\ln k/dN = 3x^2$, allows one to show from the definition of $x$ that 
\ben
x^2 = \frac{1}{12\pi \tilde G}\frac{1}{\tilde\phi^2}.
\label{e:xPhi}
\een
For a polynomial potential at a fixed point of the dimensionless couplings,
\ben
\nuRG = 4 - {\tilde\phi}\frac{\tilde V'}{\tilde V}.
\label{e:nuRGV}
\een
Hence, at a scaling simultaneous fixed point,
\ben
z = \pm \sqrt{\frac{3}{2}} x (4 - \nuRG)
\label{e:zSSFP}
\een
One may use this equation (taking the negative root), and $d\ln k/dN = 3x^2$ again, to show that the constancy of $\nuRG$ with $N$ implies
\ben
\frac{\pa \nuRG}{\pa \ln k} = (\nuRG - \siRG)(4 - \nuRG).
\label{e:nuRGSSFP}
\een
Substituting (\ref{e:zSSFP}) and (\ref{e:nuRGSSFP}) into the equation for $d \ln k /d N$ (\ref{eq:TeqnRG}), and recalling that $\etaRG = -2$ at the fixed point, one can verify that it is indeed satisfied.  Hence at a scaling simultaneous fixed point we may replace the complicated equation (\ref{eq:TeqnRG}) with the simpler $
{d \ln k}/{dN} = 3x^2.
$

We may use these equations to find an equation for the fixed point value of $\tilde\phi$, and hence the value of $\nuRG$ at the fixed point. Recalling that $x^2 = (1- 2/\nuRG)$, we find from (\ref{e:xPhi}) and (\ref{e:nuRGV}) that 
\ben
\frac{2 {\tilde V} - \tilde\phi{\tilde V'}}{4 {\tilde V} - \tilde\phi{\tilde V'}} = \frac{1}{12\pi\tilde G\tilde\phi^2}.
\label{e:GenPhiEqn}
\een
In the next two sub-sections we show that this equation reproduces those found in Section \ref{NGFP}
in the specific cases studied in this paper (monomial and quartic potential) in the fixed point regime.

\subsubsection{Monomial potential}

For the case of the monomial potential studied in Section \ref{NGFPmon}, one finds that $\nuRG$ is independent of the field, and one can directly show that, at a simulataneous fixed point, 
\begin{equation}
x=\pm\left(\frac{2-n}{4-n}\right)^{\frac{1}{2}}    \ ;\qquad y=\left(\frac{2}{4-n}\right)^{\frac{1}{2}}\ ;\qquad z=-n\,\sqrt{\frac{3}{2}\left(1+\frac{\eta_{RG}}{\nu_{RG}}\right)}=-\sqrt{\frac{3}{2}}n\, x\ .
\end{equation}
It is easy to show that 
\ben
\tilde\phi^2 = \frac{1}{12\pi\tilde G} \frac{4-n}{2-n}
\een
consistent with (\ref{e:ChiEqn}) and (\ref{e:VarPhiEqn}).

\subsubsection{Quartic potential}

With the quartic trinomial potential, one obtains from (\ref{e:GenPhiEqn})
\ben
\frac{\tilde\la_0 - \tilde\la_4\tilde\phi^4}{2\tilde\la_0 + \tilde\la_2\tilde\phi^2 } =  \frac{1}{12\pi\tilde G\tilde\phi^2},
\een
reproducing (\ref{phiequation}) as expected.

From the expressions for $\nu_{RG}$ and $\sigma_{RG}$ given in the Appendix, and the definition of $z$, we obtain
\begin{equation}
x =\pm\sqrt{\frac{\tilde\la_0 - \tilde\la_4\tilde\phi^4}{2\tilde\la_0 + \tilde\la_2\tilde\phi^2 }}
 ;\quad 
 y = \sqrt{\frac{\tilde\la_0+  \tilde\la_2\tilde\phi^2+ \tilde\la_4\tilde\phi^4}{2\tilde\la_0 + \tilde\la_2\tilde\phi^2 }}
 ; \quad z=\frac{{\tilde\phi}}{\sqrt{2\pi \tilde G}}\frac{{\tilde\lambda}_2+2{\tilde\lambda}_4{\tilde\phi}^2}
{{\tilde\lambda}_0+{\tilde\lambda}_2{\tilde\phi}^2+{\tilde\lambda}_4{\tilde\phi}^4}\ .
\end{equation}

\subsection{Quasi-classical regime}

In the quasi-classical regime the dimensionless couplings tend to zero and $\etaRG$, $\nuRG$ and $\siRG$ are also small, while $\alpha_{RG}$ does not vanish in this limit. Hence the equations (\ref{eq:XeqnRG},\ref{eq:YeqnRG},\ref{eq:ZeqnRG}) revert to their classical form. In this case 
there is a fixed point with $x=\pm1$, $y=0$, and $\eta-z^2=0$, near which trajectories emerge towards another 
at $x=0$, $y=1$, and $z=0$ (as $N$ gets more negative). Trajectories passing near this second fixed point are drawn towards the slow-roll inflationary line $x = -z/\sqrt{6}$.
Direct evaluation for field values larger than $\phi_{\rm cl}$ (so that the potential can be treated as a monomial with $n=4$) shows that for $\dot{\f}\!\simeq-A\f$ with $A\ll 1$ the phase space variables behave as 
\begin{equation}
 x_{\textrm{late}}\simeq -\sqrt{\frac{8}{3}}\frac{1}{\kappa\phi},\qquad y_{\textrm{late}}\simeq 1,\qquad z_{\textrm{late}}\simeq\frac{4}{\kappa\phi}.
\end{equation}
Using the scale identification (\ref{kvsf}) we see that 
\begin{equation}
 \frac{d\log k}{dN}\simeq \frac{4}{\ka^2\phi^2} = \frac{3}{2} x_{\textrm{late}}^2,
\end{equation}
which is clearly consistent with the general formula for $x$ (\ref{e:xPhi}) and the slow-roll condition.
Note that $k$ is not proportional to the Hubble parameter $H$ in this era: $H$ is proportional to $\phi^2$, and hence $H \propto k^2$.

\section{Cosmological fluctuations}
\label{seccomfluc}

We now want to derive an estimate for some of the CMB observables with respect to our results. To do so, we recall the fact, already stressed in~\cite{Contillo:2010ju}, that the Wilsonian RG improvement is basically an averaging procedure over a volume of radius $k^{-1}$. This means that the field fluctuations of momentum $p\gtrsim k$ should not be be affected by the variation of the coupling constants.  Furthermore, the improvement preserves the form of the classical equations of motion, which means that the two crucial ingredients in the standard calculation, the quantum mode functions and the conservation of the curvature perturbation for super-horizon modes, should be unaffected.  Hence the usual formalism of evolution of perturbations (see for example~\cite{Stewart:1993bc}) should be applicable \cite{Mukhanov:1990me,Lyth:2009zz,Baumann:2009ds}. 

\subsection{Generation of scalar perturbations}

A convenient approach is to work in the comoving gauge $T^{0i} = 0$, where the scalar field and the RG scale are functions of time only (see the end of Section \ref{s:Conservation}).  Considering scalar perturbations only, the spatial part of the metric may be written 
\ben
g_{ij} = a^2(t) e^{-2\R}\de_{ij},
\een
where $\R$ is the so-called curvature perturbation, related to the curvature of spatial sections through the three-dimensional Laplacian $\laplace$:
\ben
R^{(3)} = 4 \laplace \R.
\een
A lengthy calculation \cite{Mukhanov:1990me, Kodama:1985bj} shows that the action becomes, to quadratic order
\ben
S^{(2)}_\R = \frac12 \int d^4 x 
 \left( (v')^2 - (\partial_i v)^2 + \frac{\th''}{\th}v^2\right),
\een
where a prime denotes the derivative with respect to conformal time $\tau$, 
\ben
\theta = a \frac{\dot \phi}{H}, \quad \textrm{and} \quad v = \th \R.
\een
Note that the $k$-dependent parameters $G$ and $\Lambda$ do not appear explicitly: the dependence on the RG scale is implicit through the solutions for $\phi$ and $H$, which are different when the time-dependence of $k$ is taken into account.

To see what difference the time-dependence makes, let us 
write
$\th = \sqrt{6}a \mpl x$.  The equation for the mode function 
$v_\bp$ with wave number $\bp$ is
\ben 
v''_\bp + \left(k^2 - \frac{\th''}{\th}\right) v_\bp = 0.
\een
At a cosmological fixed point of the RG-improved dynamical equations (see Section \ref{s:CosFixPoi}) $x$ is a constant, $\mpl \propto k $, and hence $\th \propto a(\tau)k(\tau)$.
With $a \propto t^\al$ and $k \propto 1/t$, it is easy to show that $\th \propto 1/\tau$, and hence the equation for the mode function is
\ben \label{e:ModFun}
v''_\bp + \left(k^2 - \frac{2}{\tau^2}\right) v_\bp = 0, \een
where $p = |\bp|$.  
This is exactly the same as the equation for scalar mode functions in a de Sitter background, even though the background is in fact power-law inflation. It is the altered time dependence of $\dot\phi/H$ which causes this difference, although our trick of rewriting $\th$ in terms of $x$ made it look as if it came from the explicit differentiation of $\mpl$.

The solution to the mode function equation (\ref{e:ModFun}) with the correct boundary condition as $\tau \to - \infty$ is 
\ben
v_\bp =  \frac{p\tau - i}{p\tau}e^{-ip\tau}.
\een
Hence, as $\tau \to 0$ from below, corresponding to late times during the inflating epoch,
\ben
{|\R_\bp|^2} \to \frac{1}{(\th k \tau)^2}.
\een
In the standard semiclassical calculation with constant $\mpl$ and a de Sitter background, $a\tau = 1/H$, and the  formula 
\ben
\PS_{\R ,\rm cl}(p) =  \frac{1}{4\pi^2} \frac{1}{(\th \tau)^2} =\frac{1}{24\pi^2x^2} \frac{H^2}{\mpl^2} = \frac{H^2}{\dot\phi^2}\frac{H^2}{4\pi^2}
\een
follows.  However, here we have $a \propto t^\al$, and  
\ben
a\tau = \frac{\al-1}{\al H}.
\een
Noting that $\al = 1/(3x^2)$, the power spectrum of the curvature perturbation becomes
\ben
\PS_\R(p)  = \frac{1}{24\pi^2}\frac{(1- 3x^2)^2}{x^2}\frac{H^2}{\mpl^2}.
\een

\subsection{Conservation of the comoving curvature perturbation}

We now outline the proof that the conservation of the comoving curvature perturbation $\R$ outside the horizon also holds in the RG-improved cosmology with running $G$.  The standard derivation starts with the Einstein equations and the conservation of energy-momentum.  We denote the density, pressure and four-velocity of the stress-energy as $\rho$, $P$ and $u^\mu$ respectively, and define a space-dependent Hubble parameter $H(t,\bx) = \frac13 \nabla_i u^i$. The Einstein, Raychaudhuri, Euler, and continuity equations may then be written
\bea
H^2 & = & \frac{8\pi G}{3} \rho + \frac23 \laplace \R \label{e:FriEqu} \\
\dot H + H^2 & = & -\frac{4\pi G}{3}(\rho + P) + \frac13 \nabla_i a^i \label{e:RayEqu} \\
{a^i } & = & - \frac{\nabla^i P}{\rho + P} \label{e:EulEqu} \\
\frac{d\rho}{d\tpr} & = & -3H(\rho + P), \label{e:ConEqu} 
\eea
where $\tpr$ is the proper time and $a^i = d u^i/d\tpr$ is the proper acceleration.

We define the  density perturbation $\delta \rho_\bp = \rho_\bp - \bar\rho$, where $\bar\rho$ is the background density, and the gravitational potential $\Phi_\bp$ by  
\ben
-\frac23 \left( \frac{k}{aH} \right)^2 \Phi_\bp = \frac{\de \rho_\bp}{\bar\rho}.
\een
The standard analysis proceeds by differentiating the Friedmann equation (\ref{e:FriEqu}) with respect to proper time, and a lengthy rearrangement produces
\bea
\frac{1}{a^p}\frac{1}{H} \frac{d}{d t} \left( a^p \Phi_\bp \right) & = & - \frac32 (1+w) \R_\bp \\
\frac{1}{H} \frac{d}{d t} \R_\bp  & = & \frac23 \frac{c_s^2}{1+w} \left( \frac{k}{aH} \right)^2 \Phi_\bp,
\eea
where $p = (5+3w)/2$, $w = P/\rho$, and $c_s^2 = dP/d\rho$.  For long-wavelengths ($k \ll aH$) the equations may be written  
\ben
\frac{\dot\R_\bp}{H} = - \frac{2c_s^2}{5+3w} \left( \frac{k}{aH} \right)^2 \R_\bp,
\een
demonstrating that $\dot\R_\bp \to 0$ for $k/aH \to 0$, i.e.\ that the comoving curvature perturbation is constant outside the horizon.

In the restricted RG improvement, the Einstein equations and the conservation of stress-energy continue to hold. The only place where the time-dependence of the parameters has the potential to affect the proof is when differentiating the Friedmann equation with respect to time: \ben
2H \dot H = \frac{8\pi G}{3} \frac{\partial\rho}{\partial t} + \frac23 \laplace \dot \R + 8\pi \dot k \frac{\partial}{\partial \ln k}(G\rho). 
\een
However, the last term vanishes as part of the consistency condition for the RG improvement to maintain the Bianchi identity (\ref{constraint}). Hence the proof continues as for the classical case, and the comoving curvature perturbation is indeed conserved outside the horizon in this RG-improved framework.

\subsection{Tensor perturbations}

For tensor perturbations, we write the spatial part of the metric tensor as 
\ben
g_{ij} = a^2(t) (\de_{ij} + 2E_{ij}),
\een
where $E_{ij}$ is symmetric, transverse ($\pa_i E_{ij}=0$) and traceless ($E_{II} = 0$).  Substitution into the Einstein-Hilbert action gives 
\ben
S^{(2)}_t = \frac12 \int d^4 x Z^2 
 \left( (E'_{ij})^2 - (\partial_l E_{ij})^2\right),
\een
where $Z = a\mpl$.  Writing $h_{ij} = Z E_{ij}$, and decomposing $h_{ij}$ into its polarisation components,
\ben
h_{ij}  = h_+ e^+_{ij} + h_\times e^{\times}_{ij}, 
\een
we find
\ben
S^{(2)}_t = \frac12\sum_A \int d^4 x  
 \left( (h'_{A})^2 - (\partial_l h_{A})^2 + \frac{Z''}{Z}(h_{A})^2\right),
\een
with $A = +,\times$. Hence the equations for the mode functions are 
\ben
h''_{A,\bp} + \left(p^2 - \frac{(\nu^2-1/4)}{\tau^2}\right)h_{A,\bp} = 0.
\een
where $p = |\bp|$ and 
\ben
\nu = \frac32 + \frac{1}{\al-1}.
\een
This is the standard equation for mode functions in a power-law inflation background, for which the solution with the correct boundary conditions can be expressed in terms of a Hankel function \cite{Lyth:2009zz}
\ben
h_{A,\bp} =  \frac12 \sqrt{\pi}e^{i(\nu + \half)\frac{\pi}{2}}(-\tau)^\half H_\nu^{(1)}(-p\tau).
\een
As $\tau \to 0$ from below, corresponding to late times during the inflating epoch,
\ben
h_{A,\bp} \to e^{i(\nu - \half)\frac{\pi}{2}}2^{\nu - \frac32} \frac{\Ga(\nu)}{\Ga(\frac32)} \frac{1}{\sqrt{2p}}(-p\tau)^{\half-\nu}.
\een
Hence 
\ben
{|E_{ij,\bp}|^2} = \sum_A\frac{|h_{A,\bp}|^2}{Z ^2} \ \to  2\frac{1}{a^2\tau^2\mpl^2}\left(\frac{\Ga(\nu)}{\Ga(\frac32)}\right)^2 \frac{1}{2p^3}\left(\frac{-p\tau}{2}\right)^{3-2\nu}  .
\een
We have $1/(a\tau) = \al H/(1-\al)$, so the tensor power spectrum is given by 
\ben
\PS_{t }(p) \equiv \frac{p^3}{2\pi^2}{|E_{ij,\bp}|^2} = \frac{2^{2\nu-3}}{2\pi^2}\left(\frac{\Ga(\nu)}{\Ga(\frac32)}\right)^2 \left.{(1- 3x^2)^{2(1+\nu)}}\frac{H^2}{\mpl^2}\right.
\left(-p\tau\right)^{3-2\nu}  ,
\een
The form is identical to the standard formula for power-law inflation. However, at the combined RG and cosmological fixed point $H/\mpl$ is a constant, rather than decreasing as $\tau^{-1/(\al-1)}$.  As $3 - 2\nu = - 2/(\al -1)$ is negative, this means that the overall normalisation of the tensor power spectrum increases as inflation proceeds, and vanishes in the infinite past at the fixed point.

Note that the infrared divergence of  power-law inflation is also present: the fluctuations diverge as $p \to 0 $.  In the standard picture it is supposed that inflation had a beginning in a finite region of the universe, whose size acts as an IR cut-off.  In our case of an emergence from a UV simultaneous cosmological and RG fixed point one can impose an IR cut-off with a toroidal compactification.  

\subsection{Amplitude and tilt of scalar power spectrum}

We can now estimate the amplitude and tilt of the scalar power spectrum.
We see that the tilt is exactly zero, as the power spectrum is independent of $p$ at the fixed point
\ben
n_s - 1 = \frac{d \ln \PS_{\R} }{d\ln p} = 0.
\een
The amplitude of the scalar power spectrum is 
\ben
\PS_{\R} = \frac{1}{\pi} \frac{\tilde G \al^3}{\chi^2}\left(1 - \frac{1}{\al} \right)^2.
\een
For the symmetric quartic potential, we may substitute from (\ref{e:AlpChiPhi}) to find (assuming $\al$ is large)
\ben
\PS_{\R} \simeq \frac{32}{3} {\pi} \tilde G^3 \tilde \phi^2 (2\tilde\la_0 + \tilde \la_2\tilde\phi^2).
\een
Asymptotically safe fixed points in the truncations studied to date \cite{Reuter:1996cp,Dou:1997fg,Souma:1999at,Lauscher:2001ya,Lauscher:2002sq,Reuter:2001ag,Litim:2003vp,Percacci:2003jz,Percacci:2005wu,Litim:2006dx,Niedermaier:2006wt,Niedermaier:2006ns,Percacci:2007sz,Litim:2008tt,Codello:2006in,Codello:2007bd,Codello:2008vh,Narain:2009fy,Narain:2009gb,Benedetti:2009gn,Machado:2007ea} have $\tilde G = {\cal O}(1)$.  
Thus we see that to arrange for a small power spectrum, the fixed point couplings $\tilde\lambda_n$ ($n = 0,2,4$) must be tuned such that either $\tilde\phi^2$ or  $2\tilde\la_0 + \tilde \la_2\tilde\phi^2$ are small. The first possibility is excluded because $\tilde\phi^2$ has to be of the same order as $\alpha$ by Eq. (\ref{mufromalpha}), which however is required to be large. The second possibility implies simultaneous vanishing of the left and right hand sides of Eq. (\ref{phiequation}). The left hand side of Eq. (\ref{phiequation}) implies then $\tilde\phi^2\simeq\sqrt{r_0}$, and the right hand side gives $\tilde\phi^2\simeq -2r_0/r_2$. This means that in order to obtain small curvature perturbations we must stay close to the line  $r_2=-2\sqrt{r_0}$ in the parameter space. The deviation from this line can be parametrized by the smallness parameter $\delta$ defined by
\begin{equation}\label{r2para}
r_2=-2\sqrt{r_0}+\frac{\delta}{\tilde G}
\end{equation}
and the curvature perturbations become
\ben
\PS_{\R} \simeq \frac{32}{3} {\pi} {\tilde G}^2 \tilde\lambda_0\delta=\frac{4}{3}\tilde{G}{\tilde\Lambda}\,\delta \ .
\een
With the parametrisation (\ref{r2para}) it also follows directly that 
\begin{equation}
\alpha\simeq 4\pi\tilde{G}\left(\sqrt{r_0}+\frac{\delta}{2\tilde G}\right)\ .
\end{equation}
Note that this entails that $\alpha$ diverges in the limit $\tilde\lambda_4\rightarrow 0$, corresponding to the Gaussian matter fixed point.
It is easy to see that the condition $r_2=-2\sqrt{r_0}$ results in the dimensionless potential being zero at its minimum, ${\tilde\phi}^2=-2r_0/r_2$.

\subsection{Amplitude and tilt of tensor power spectrum}

Assuming that $\al$ is large, we have
\ben
\PS_{t }(p) \simeq \frac{1}{2\pi^2} \frac{H^2}{\mpl^2}(-p\tau)^{n_T}
\een
where $n_T$ is the tensor tilt, given  by 
\ben
n_T = -\frac{2}{\al - 1}.
\een
Hence using (\ref{e:AlpChiPhi}), 
\begin{equation}
\PS_{t }(p) \simeq \frac{4\tilde G}{\pi} \frac{\al^2}{\chi^2}\left(-p\tau\right)^{n_T}
= \frac{32}{3}\tilde G^2 (2\tilde\la_0 + \tilde \la_2\tilde\phi^2)\left(-p\tau\right)^{n_T} \
\end{equation}
which, using (\ref{r2para}), can be written as
\begin{equation}
\PS_{t }(p)\simeq\frac{16}{3}\sqrt{\frac{\tilde\Lambda\tilde G{\tilde\lambda}_4}{2\pi}}\delta \left(-p\tau\right)^{n_T} \ .
\end{equation}
Thus, the tensor-to-scalar ratio is
\begin{equation}
r = \frac{\PS_{t }(p)}{\PS_{\R }(p)}\simeq \sqrt{\frac{8{\tilde\lambda}_4}{\pi\tilde\Lambda\tilde G}}\left(-p\tau\right)^{n_T} \ .
\end{equation}
The dimensionless combination of gravitational couplings $\tilde G\tilde \Lambda$ is thought to be close to order unity at the UV fixed point (see e.g.\ \cite{Reuter:2007rv}), while  $\tilde\lambda_4$  vanishes if the UV behaviour of the scalar field is controlled by a Gaussian matter fixed point. Non-zero values of $r$ therefore require non-trivial UV behaviour in the scalar sector.

\section{Conclusions}

In this paper we have provided a description of the early history of the universe, where the matter content is described by a scalar field, and gravitation by the Einstein-Hilbert action, whose coupling constants vary according to the renormalisation group equations.  The RG scale is assumed to be time-dependent, and varies in such a way as to preserve the form of the classical equations of motion.
We assume that there is a UV fixed point at which the dimensionless couplings approach constant values.

Expanding solutions, in the form of power law, were found in the  fixed point regime for the field potentials of monomial and polynomial form. 
We  found that the simplest form of monomial potentials does not provide accelerated expansion in the RG fixed point regime.

We considered a potential including all even non-singular and perturbatively renormalisable interaction terms and found accelerated expansion in the form of power-law inflation for a very wide range of model parameters. The results were classified as cosmological fixed points of an autonomous phase-space portrait.

We also studied the transition from the UV regime to classicality. We found that the fully classical regime, in which the coupling constants reach their threshold values and freeze, is an attractor solution for the system and an attracting fixed point in the cosmological phase-space. This is an important result, because it ensures that the theory has a safe late-time regime that can be matched with the standard model of cosmology below the Planck-scale.  In this regime, the consistency of the improved Einstein equations forces the RG scale to be identified with the scalar field strength.

Finally, we calculated the scalar and tensor perturbations induced by the fluctuations in the scalar and gravitational fields during the inflationary phase, taking into account the time-dependence of the gravitational coupling $G$.  The standard formulae are recovered, but the  time-dependence of $G$ means that the amplitude and tilt of the power spectra are unusual.  The scalar power spectrum is exactly scale-invariant, while the tilt of the tensor power spectrum is as expected for power-law inflation.  The amplitudes are generically of ${\cal O}$(1), but can be small if the fixed-point values of the dimensionless couplings in the scalar potential obey certain relations which tune the dimensionless potential to zero.  Indeed, as the scale of inflationary perturbations is set by the product $GH^2$, and near a fixed point the gravitational constant $G \sim 1/k^2$, it is clear that in order to generate small inflationary peturbations we must have $H \ll k$. The RG scale is instead the same order of magnitude as the scalar field strength.

Hence it is possible that under the special circumstances detailed above the cosmological fluctuations we observe are generated near a simultaneous cosmological and RG fixed point. The attraction of this picture is that the fluctuations of the fields are under control all the way to the initial singularity, as long as the universe has finite comoving size. If a suitable fixed point is found, it would therefore constitute a viable UV completion of the inflationary paradigm.

\begin{acknowledgments}
We would like to thank Daniel Litim, Roberto Percacci, Ninfa Radicella and Martin Reuter for useful discussions. This work was supported by the Science and Technology Research Council [grant number ST/G000573/1]. 
\end{acknowledgments}

\section{Appendix}

For a monomial potential of type $V(\phi)=\l_n\phi^n$, one obtains the RG parameter
\begin{equation}
\nu_{RG}=\sigma_{RG}=4-n+\frac{\beta_{{\tilde g}_n}}{{\tilde g}_n}\ ;\qquad \alpha_{RG}=1-\frac{n}{2}\ .
\end{equation}
For the quartic potential in four dimensions, one obtains
\begin{eqnarray}
\nu_{RG}&=&\frac{1}{\tilde V}\left((\b_{\lt_0}+4\lt_0)+(\b_{\lt_2}+2\lt_2){\tilde \f}^2+\b_{\lt_4}{\tilde\f}^4\right)\\
\sigma_{RG}&=&\frac{2\, \tilde\phi}{{\tilde V}'}\left((\b_{\lt_2}+2\lt_2)+2\b_{\lt_4}{\tilde\f}^2\right)\;.
\end{eqnarray}
At an RG fixed point, these expressions simplify to 
\begin{eqnarray}
\nu_{RG}&=&\frac{2\left(2{\tilde\lambda}_0+{\tilde\lambda}_2{\tilde\phi}^2\right)}
{{\tilde\lambda}_0+{\tilde\lambda}_2{\tilde\phi}^2+{\tilde\lambda}_4{\tilde\phi}^4}\\
\sigma_{RG}&=&\frac{2{\tilde\lambda}_2}{{\tilde\lambda}_2+2{\tilde\lambda}_4{\tilde\phi}^2}\\
\alpha_{RG}&=&\frac{2{\tilde\lambda}_0}{2{\tilde\lambda}_0+2{\tilde\lambda}_2{\tilde\phi}^2}
\end{eqnarray}
where the solutions from Section \ref{secquartic} have been inserted. 

\bibliography{AS_references}
\end{document}